\documentclass[aps,prl,twocolumn,showpacs]{revtex4}
\usepackage{amssymb}

\usepackage{graphicx}

\begin{document}

%draft

\title{Nodeless superconductivity in K$_x$Fe$_{2-y}$Se$_2$ single crystals revealed by low temperature specific heat}

\author{Bin Zeng$^1$, Bing Shen$^1$, Gengfu Chen$^2$, Jianbao He$^2$, Duming Wang$^2$, Chunhong Li$^1$,  and Hai-Hu Wen$^{1,3}$}\email{hhwen@nju.edu.cn}

\affiliation{$^1$ Institute of Physics and Beijing National
Laboratory for Condensed Matter Physics, Chinese Academy of
Sciences, P.O. Box 603, Beijing 100190, People's Republic of China}

\affiliation{$^2$ Department of Physics, Renmin University of China,
Beijing 100872, China}

\affiliation{$^3$ National Laboratory of Solid State Microstructures
and Department of Physics, Nanjing University, Nanjing 210093,
China}

\begin{abstract}
Low temperature specific heat (SH) has been measured in
K$_x$Fe$_{2-y}$Se$_2$ single crystals with T$_c$ = 32 K. The SH
anomaly associated with the superconducting transition is moderate
and sharp yielding a value of $\Delta C/T \mid_{T_c}$ = 11.6 $\pm$
1.0 $mJ/mol K^2$. The residual SH coefficient $\gamma(0)$ in the
superconducting state at T$\rightarrow$ 0 is very small with a value
of about 0.39 $mJ/mol K^2$. The magnetic field induced enhancement
of the low-T SH exhibits a rough linear feature indicating a
nodeless gap. This is further supported by the scaling based on the
s-wave approach of the low-T data at different magnetic fields. A
rough estimate tells that the normal state SH coefficient $\gamma_n$
is about 6 $\pm$ 0.5 $mJ/mol K^2$ leading to $\Delta
C/\gamma_nT\mid_{T_c}$ = 1.93 and placing this new superconductor in
the strong coupling region.
\end{abstract}

 \maketitle

The discovery of high temperature superconductivity in iron
pnictides has opened a new era towards the investigation on the
novel superconducting mechanism.\cite{Kamihara2008} One of the key
issues here is about the superconducting pairing mechanism.
Experimentally it was found that the superconductivity is at the
vicinity of a long range antiferromagnetic (AF) order\cite{DaiPC},
the superconducting transition temperature is getting higher when
this AF order is suppressed. It was also further proved that the AF
spin fluctuation\cite{ImaiNMR} and the multi-band
effect\cite{FangLei} are two key factors for driving the system into
superconductive. Theoretically several different pairing symmetries
are anticipated. It was suggested that the pairing may be
established via inter-pocket scattering of electrons between the
hole pockets (around $\Gamma$ point) and electron pockets (around M
point), leading to the pairing manner of an isotropic gap on each
pocket but with opposite signs between them (the so-called
S$^\pm$).\cite{Mazin,Kuroki,LeeDH,LiJX} The pairing picture based on
the super-exchange of local moment was also proposed, which in
principle could also lead to the S$^\pm$\cite{HuJP,SiQM}, leaving
the others (d-wave or full-gapped S$^{++}$) as perspectives with low
possibility. However, the S$^{++}$ pairing manner is specially
winning the vote when the orbital fluctuation plays the important
role, as argued by Kotani et al.\cite{Kotani}. Clearly multi-orbits,
or the naturally formed multi-pockets are highly desirable for the
superconductivity of all these pairing models.

Recently a new Fe-based superconducting system $A_xFe_{2-y}Se_2$ (A=
alkaline metals, x$\leq$1, y$\leq0.5$) were discovered with the
transition temperature above 30 K.\cite{ChenXL} The interests to
this fascinating system are immediate because of the two major
reasons:(1) Both the band structure calculations\cite{BandCal,LuZY}
and the preliminary angle resolved photo-emission spectrum (ARPES)
measurements\cite{FengDL,DingH,ZhouXJ} indicate that the band near
the $\Gamma$-point seems diving far below the Fermi energy, leading
to the absence of the hole pockets which are widely expected for the
FeAs-based systems. A consequence of these results is to question
the importance of the inter hole-electron pocket scattering for the
superconducting pairing. (2) The superconducting state seems
occurring via a transmutation from an insulating ordered state of
Fe-vacancies.\cite{FangMH} The question raised here is whether this
insulating state originates from the Mottness, like in the
cuprate,\cite{SiQM2} or the band gap due to the reconstruction of
the electronic structure when the Fe vacancies are present. Many
different kind of pairing symmetries are proposed, such as nodeless
d-wave,\cite{LeeDH2} S$^{++}$ or S$^\pm$, all are satisfying with
the basic structures. The experimental evidences about the
superconducting gaps so far are quite rare. The ARPES measurements
indicate isotropic gaps on the four electron pockets with a rather
large gap value ($\sim$ 8-15 meV). The conclusions drawn from NMR
measurements seem controversial.\cite{YuWQ,Nakai} In this paper, we
present the first set of data of low temperature specific heat (SH)
measurements. Our detailed analysis indicates a nodeless gap with a
strong coupling strength in this new superconducting system.

The K$_x$Fe$_{2-y}$Se$_2$ single crystals were synthesized by the
flux-growth method\cite{ChenGFKFeSe}. The typical dimension of the
samples for specific heat measurements was 2$\times$2$\times$0.5
mm$^3$. The SH measurements were done with the thermal relaxation
method on the Quantum Design instrument physical property
measurement system (PPMS) with the temperature down to 2 K and
magnetic field up to 9 T. The magnetic field effect on the bare SH
measuring chip (including the four thermal conducting wires) of PPMS
from Quantum Design was calibrated prior to the measurements on the
samples, in order to remove the pseudomorphism. This becomes very
essential since the contribution of the electronic SH is quite small
compared to the phonon part in this particular system. The dc
magnetization measurements were done with a superconducting quantum
interference device (Quantum Design, SQUID).

\begin{figure}
\includegraphics[width=8cm]{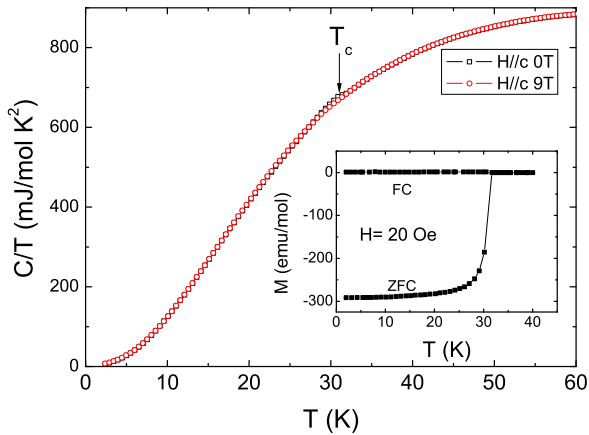}
\caption {(color online) Raw data of the temperature dependence of
specific heat for the K$_x$Fe$_{2-y}$Se$_2$ at 0T and 9T. The inset
shows the temperature dependence of the dc magnetization for the
same sample measured under a magnetic field of 20 Oe. } \label{fig1}
\end{figure}

In Fig.1, we show the temperature dependence of specific heat and dc
magnetization for the sample. The sharp transition in the
magnetization and the large magnetic screening signal indicate the
good quality of the sample. In the SH data at zero field, one can
see a small feature of heat capacity anomaly at T$_c$. We will show
that this SH anomaly is actually rather sharp with a reasonable
magnitude.

The SH data for the sample in low temperature region are plotted as
$C/T$ vs $T^2$ in the inset of Fig.2. No Schottky anomaly was
detected. The data below about 8 K can be fitted by
\begin{equation}
C(T,H)=\gamma(H) T+\beta T^3+\eta T^5,\label{eq:1}
\end{equation}
where $\gamma(H)T$ is the residual SH coefficient in the magnetic
field $H$, $\beta T^3+\eta T^5$ is the phonon part of heat capacity.
Normally it is unnecessary to consider the last term $\eta T^5$ in
the low temperature region, while this seems not the case for the
present sample. One can see a slight upturn curvature in the low-T
SH data $C/T$ vs. $T^2$. This is understood because of the
relatively low Debye temperature of the sample, as discussed below.

For the heat capacity of the sample, the phonon contribution should
be identical in zero field and in magnetic field. Hence, the
parameters $\beta$ and $\eta$ should be the same for 0T and 9T. This
is a constraint on the fitting process of the data. By fitting the
SH data in 0T and 9T using Eq.(1), we obtained $\gamma(0)$ $\approx$
0.394 mJ/mol K$^2$, $\gamma(9T)$ $\approx$ 1.4 mJ/mol K$^2$, $\beta$
$\approx$ 1.018 mJ/mol K$^4$ and $\eta$ $\approx$ 0.003 mJ/mol
K$^6$. Using the obtained value of $\beta$ and the relation
$\Theta_D = (12\pi^4k_BN_AZ/5\beta)^{1/3}$, where $N_A$ = 6.02
$\times$ 10$^{23}$ mol$^{-1}$ is the Avogadro constant, Z = 5 is the
number of atoms in one unit cell, we get the Debye temperature
$\Theta_D$ $\approx$ 212 K, which is relatively small, compared to
other FeAs-based superconductors\cite{Mugang1111,Mugang122K}.

\begin{figure}
\includegraphics[width=8cm]{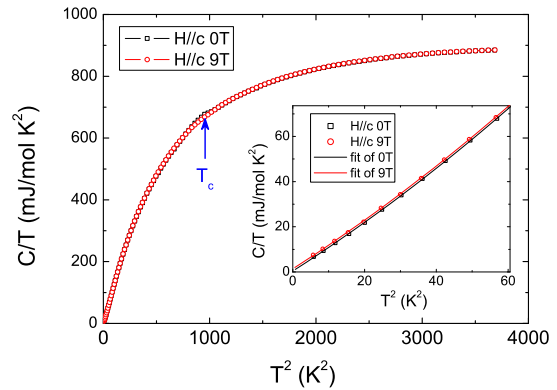}
\caption {(color online) The main panel is the SH data plotted as
$C/T$ vs $T^2$. The inset is the enlarged view of the data together
with the fit (see text) in low temperature region, and no Schottky
anomaly was detected. } \label{fig2}
\end{figure}

Fig.3(a) shows the enlarged view of the SH data near the transition
temperature plotted as $C/T$ vs $T$. One can see a clear SH anomaly
at $T_c$ in 0T, and when a magnetic field is applied, the SH anomaly
was weakened and shifted to lower temperatures. Fig.3(b) shows the
difference of the SH data between 0T and 9T. The SH anomaly at $T_c$
is more obvious. We can evaluate the height of the SH anomaly
$\Delta C/T|_{Tc}$ near $T_c$ from the difference of $C/T$ at 0T and
9T. The estimated anomaly $\Delta C/T|_{Tc}$ is about 11.6 $\pm$ 1
$mJ/mol K^2$. For the optimally doped (Ba,K)Fe$_2$As$_2$ and
Ba(Fe,Co)$_2$As$_2$, the SH anomaly $\Delta C/T|_{Tc}$ are about 98
$mJ/mol K^2$ and 28.6 $mJ/mol K^2$,
respectively\cite{Mugang122K,Canfield,Ronning,Keimer,Mugang122Co}.
The SH anomaly for K$_x$Fe$_{2-y}$Se$_2$ is also smaller than other
FeAs-based superconductors. The SH anomaly looks rather sharp (it
starts at about 32.9 K and ends at 30.9 K). We must emphasize that
to use the data measured at 9 T as the background to deduce the SH
anomaly is reasonable. As we address below that a magnetic field of
9 T should have lowered down the transition temperature of about 5-6
K (with the upper critical field $H_{c2(0)}^c \approx 48T$), being
much larger than the width of the SH anomaly. The rather sharp SH
anomaly is very different from that in the underdoped cuprates in
which a long-tail of electronic SH was observed far into the normal
state.\cite{WenHHPRL} This was interpreted as the fluctuating
superconductivity. This may suggest that, although having a low
superfluid density and relatively higher anisotropy,\cite{LiCH} the
superconducting transition in A$_x$Fe$_{2-y}$Se$_2$ superconductors
can still be described quite well by the critical mean field theory
without the necessity of categorizing it into the strong critical
fluctuation.

\begin{figure}
\includegraphics[width=7cm]{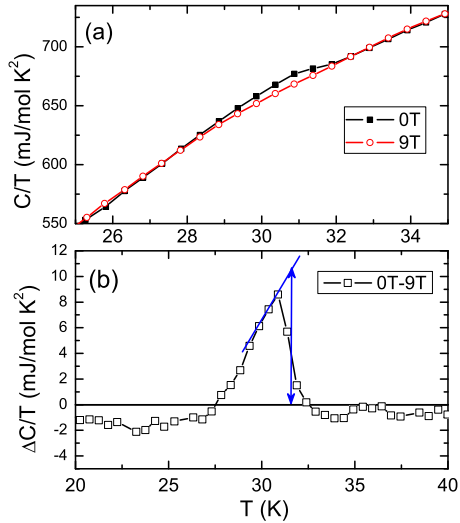}
\caption {(color online) The upper panel is the enlarged view of the
SH data near the transition temperature plotted as $C/T$ vs $T$. The
lower panel shows the difference of the SH data between 0T and 9T.
The blue solid lines here are just guides to the eyes for showing
how we determine the height of the SH anomaly.} \label{fig3}
\end{figure}

Fig.4(a) shows the SH data plotted as $C/T$ vs $T^2$ in low
temperature region under various magnetic fields. One can see that
the magnetic field enhances the SH coefficient progressively,
indicating the generating the quasiparticle density of sates.
Fig.4(b) presents the difference of the SH data between a typical
magnetic field and zero field and the dashed lines are the linear
fit of the difference in the temperature region of 2.4 K to 4 K. The
slight bending down of the data here below about 2.4 K at fields of
5, 7, 9 T was recognized as due to the unsuccessful removing of the
magnetic field effect on the SH measuring chip. And this would have
a more obvious effect on high magnetic field data, and especially
the low-T region. As shown in Fig.4(b), the difference of SH data
for 9T and 0T is a little tilted , rather than a rough constant, so
we extracted the data at T = 3K as the field induced term of SH.
Using the linear fit (between 2.4 and 4 K) in panel (b), we can
obtain the magnetic field induced enhancement of the low-T SH, and
the extracted data at T = 3K are shown in Fig.4(d). It is clear that
the field induced term exhibits a roughly linear field dependence.
In combination with the fact that a very small residual SH term was
observed in the superconducting state at T$\rightarrow$0, we tempt
to conclude a nodeless gap.

In order to further confirm this point, we analyzed the SH data in
finite-temperature region in the mixed state. It is known that the
quasiparticle excitations in superconductors with different gap
symmetries can be obviously distinct. In s-wave superconductors, the
inner-core states dominate the quasiparticle excitations, and
consequently a simple scaling law $C_{QP}/T^3 \approx C_{core}/T^3 =
(\gamma_n/H_{c2(0)}\times(T/\sqrt H)^{-2}$ for the fully gapped
superconductors is expected, where $C_{QP}$ and $C_{core}$ are the
specific heat of the quasiparticles induced by field and that from
the vortex cores in the mixed state, respectively. The scaling
result of the field-induced term in the mixed state with the s-wave
condition is presented in Fig.4(c). One can see that all the data at
different magnetic fields can be roughly scaled to the straight blue
line, which reflects the theoretical curve $C_{cal-s} = 0.12(T/\sqrt
H)^{-2}$. Generally, this prefactor $\gamma_n/H_{c2(0)} = 0.12
mJ/(mol K^2 T)$ is consistent with the magnitude of the slope of the
line in Fig.4(d). Using the value of $H_{c2(0)}^c \approx 48
T$\cite{ChenGFKFeSe}, we estimate the value of normal-state electron
SH coefficient $\gamma_n$ to be 5.8 $mJ/mol K^2$, which is a small
value compared to other FeAs-based
superconductors\cite{Mugang122K,Canfield,Ronning,Keimer,Mugang122Co}.

\begin{figure}
\includegraphics[width=9cm]{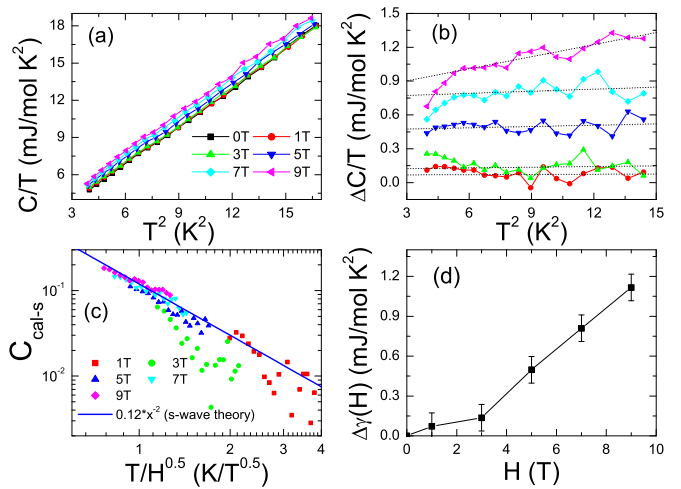}
\caption {(color online) (a) The SH data in low temperature region
under various magnetic fields. (b) The difference of the SH data
between a particular magnetic field and zero field. The dashed lines
are the linear fit of this difference between 2.4 K - 4 K. The
slight bending down of the data at 5, 7, 9 T was recognized as due
to the unsuccessful removing of the magnetic field effect on the SH
measuring chip. (c) Scaling of the data according to the s-wave
scenario (symbols) $C_{cal-s} = [C(H)-C(0)]/T^3$ vs $T/\sqrt H$, the
dashed line represents the theoretical expression. (d) Field
dependence of the field-induced term $\Delta \gamma(H) =
[C(T,H)-C(T,0)]/T$ at T = 3K.} \label{fig4}
\end{figure}

As far as we know, reliable calculated values for the normal state
DOS and thus the SH coefficient of this new superconductor are still
lacking because of the uncertainties of the structures of these
Fe-vacancies. The $\gamma_n \approx$ 6 $mJ/mol K^2$ found here will
give some hint on the band structures
 as well as understanding the ARPES data. The value $\Delta
C/\gamma_nT\mid_{T_c}$ = 1.93 clearly places the system in the
strong coupling camp, since the weak coupling BCS theory gives 1.43.
Furthermore, the very small residual SH coefficient
$\gamma(0)\approx$ 0.39 $mJ/mol K^2$ together with the s-wave
scaling excludes the nodal gaps in this system. This is consistent
with the ARPES data so far, and in contradicting with the NMR
data.\cite{Nakai} The small $\gamma(0)$ detected here excludes also
the chemical phase separation picture, since otherwise a much
significant value, arising from the non-superconducting normal
metallic regions, should be observed. However, if the system is
chemically separated into the superconducting regions and the
insulating regions which are fully gapped, this is of course
acceptable. Our results here is also against with the nodeless
d-wave picture since that kind of pairing is certainly suffered
sensitively from the impurity scattering which would give rise to a
large quasiparticle density of states detectable by specific heat.
The sharp SH anomaly found here indicates that the present system
does not have a strong critical fluctuation which appears in the
underdoped cuprates.

In summary, we measured the low temperature SH of single crystal
K$_x$Fe$_{2-y}$Se$_2$ in various magnetic fields. The SH anomaly is
observed at $T_c$ = 32K, and the height $\Delta C/T\mid_{T_c}$ is
about 11.6 $mJ/mol K^2$. From the low temperature part of the SH
data, we obtained the field induced enhancement of the low-T SH,
which exhibits a roughly linear field dependence, indicating a
nodeless gap. We also analyzed the data with the s-wave scaling law,
and found that the data roughly obey this law, indicating again an
s-wave gap. These two approaches are self-consistent each other. The
Debye temperature and the normal-state electronic SH coefficient
were also estimated, both are smaller than that in other FeAs-based
superconductors.

\begin{acknowledgments}
We appreciate the useful discussions with Hong Ding, Doug Scalapino,
and Tao Xiang. This work is supported by the NSF of China, the
Ministry of Science and Technology of China (973 projects:
2011CBA001000), and Chinese Academy of Sciences.
\end{acknowledgments}

$^{\star}$ hhwen@nju.edu.cn

\end{document}